\documentclass[english,prd,twocolumn,superscriptaddress,nofootinbib,preprintnumbers]{revtex4-1}
\usepackage[latin1]{inputenc}
\usepackage{graphicx}
\usepackage{amssymb}
\usepackage{color}
\usepackage{float}
\usepackage{amsmath}
\usepackage{amsfonts}
\usepackage{dcolumn}
\usepackage{hyperref}
\usepackage{subfigure}
\usepackage{mathtools}
\usepackage{soul,xcolor}

\newcommand{\iu}{{i\mkern1mu}}
\newcommand{\R}{\mathbb{R}}

\begin{document}

\title{Black hole stability under odd-parity perturbations in Horndeski gravity}

\author{Apratim Ganguly}
\affiliation{Department of Mathematics, Rhodes University, 6140 Grahamstown, South Africa}
\author{Radouane Gannouji}
\author{Manuel Gonzalez-Espinoza}
\author{Carlos Pizarro-Moya}
\affiliation{Instituto de F\'{\i}sica, Pontificia Universidad  Cat\'olica de Valpara\'{\i}so, Casilla 4950, Valpara\'{\i}so, Chile}

\begin{abstract}
We study the stability under linear odd-parity perturbations of static spherically symmetric black holes in Horndeski gravity. We derive the master equation for these perturbations and obtain the conditions of no-ghost and Laplacian instability. In order for the black hole solutions to be stable, we study their generalized ``Regge-Wheeler potential''. It turns out that the problem is reduced to an algebraic problem where three functions characterizing the black hole should be positive outside the horizon to prove the stability. We found that these conditions are similar to the no-ghost and Laplacian instability conditions. We apply our results to various known solutions.
\end{abstract}

\maketitle

\section{Introduction}%%%%%%%%%%%%%%%%%%%%%%%%%%%%%%%%%%%%%%%%%

Extended models of gravity became very popular in cosmology and black hole physics. In cosmology, these models are used to explain inflation, dark energy or dark matter, while hairy black hole solutions are studied in the case of compact objects. They can have a very rich phenomenology (see \cite{Berti:2015itd} for a recent review). In this paper, we focus on static black hole solutions and their stability.

In fact, since the seminal work by Regge and Wheeler \cite{Regge:1957td} followed by Zerilli's analysis \cite{Zerilli:1970se}, the theory of linear perturbations is well defined. The metric perturbations are decomposed according to their transformation properties under two-dimensional rotations. They are classified as odd (axial or vector) perturbations and even (polar or scalar) ones. The two modes give rise to a master linear differential equation similar to the Schr\"{o}dinger equation, from which, the stability can be derived, following the work by Vishveshwara \cite{Vishveshwara:1970cc}, Price \cite{Price:1971fb} and Wald \cite{Wald:1979}. This formalism has also been extended to higher dimensions where an additional mode, dubbed tensor perturbations, appears \cite{Gibbons:2002pq}. The procedure to study these solutions is also well-known. Developed initially by \cite{Kodama:2003jz,Ishibashi:2003ap}, it has been applied to various solutions (see e.g. \cite{Gleiser:2005ra,Takahashi:2010ye,Gannouji:2013eka}).

Instead of studying a particular model, we intend to give the generic conditions of stability of static black holes in Horndeski model \cite{Horndeski:1974wa}. These models which appeared for the first time in 1974 didn't attract much interest, until they were rediscovered in a different context: brane models. Extra dimensions provide an approach to modify gravity without abandoning the form of the action proposed in Einstein's general relativity, such as Lovelock polynomial \cite{Lovelock:1971yv}. From a phenomenological point of view, we can avoid the constraints coming from the standard model, by considering a brane world scenario instead of a compactified dimension, that is, we are living in a hypersurface (the brane) in a higher dimensional spacetime (the bulk). From the theoretical point of view, string theory predicts a boundary layer, a brane, on which edges of open strings stand. The possibility that we may be living in a brane generates many questions and possible solutions to numerous debated problems such as the hierarchy problem \cite{Gogberashvili:1998vx}.

Among brane models, DGP \cite{Dvali:2000hr} attracted a lot of attention even though an induced curvature term can't be motivated by higher energy theories and the self-accelerating branch is plagued by an instability \cite{Luty:2003vm,Nicolis:2004qq,Koyama:2005tx,Koyama:2005br,Charmousis:2006pn}. However, the analysis of the model continued, and an effective theory in 4D has been derived by integrating out the extra dimension, giving rise to the so-called galileon, \cite{Luty:2003vm} where the scalar field stands for the brane-bending mode (i.e., a longitudinal graviton). It has been extended with additional terms with the same symmetry in Minkowski background \cite{Nicolis:2008in}. The model has been first promoted to a covariant form \cite{Deffayet:2009wt} by abandoning the galilean symmetry and finally generalized to the most general action with a single scalar field which leads to second order differential equations \cite{Deffayet:2011gz}, keeping the theory free from the Ostrogradsky instability \cite{Woodard:2006nt}. This model has been shown to be Horndeski's original work \cite{Kobayashi:2011nu}. These models have not only attracted the attention of cosmologists but also the community of black hole physics, because it constitutes a simple field to check various no-go theorems.

In this paper, we will derive the full stability conditions of a black hole solution subjected to a linear odd-parity gravitational perturbation. This work has already been approached by other authors \cite{Kobayashi:2012kh}. 
In their paper, they have expanded the Horndeski action to second order in odd-parity perturbations and succeeded in identifying the master variable, they obtained the no-ghost and no-gradient instability conditions (see \cite{Kobayashi:2012kh} for more details). All these results have been recovered in our paper, but we have derived the correct generalized Regge-Wheeler equation\footnote{Our result has been accepted in a private communication with the authors of \cite{Kobayashi:2012kh} and it will be proved in a simple example as shown in the Appendix \ref{appenA}.} which is fundamental in the study of the stability of black holes and quasinormal modes. We have performed the full analysis of the stability condition, by expanding the action to second order of perturbations, obtain the no-ghost and Laplacian stability conditions and derive the correct generalized Regge-Wheeler potential. Due to quasinormal modes (QNM) oscillations, we discuss the stability of black holes by using the S-deformation technique \cite{Ishibashi:2003ap}. Our result reduces to previous expressions derived in the literature, such as the Regge-Wheeler potential \cite{Regge:1957td} or more recently \cite{Anabalon:2014lea,Cisterna:2015uya}, in presence of a scalar field. We find that the stability analysis is reduced to an algebraic problem where three functions characterizing the black hole should be positive outside the horizon. We finally apply this formalism to various solutions, checking at the same time that our calculations are correct, by confronting them to cases previously discussed in the literature.
                      
\section{The model}%%%%%%%%%%%%%%%%%%%%%%%%%%%%%%%%%%%%%%%%%

Following the same strategy as \cite{Kobayashi:2012kh}, we consider the most general scalar-tensor theory in four dimensions having second-order field equations, both for the metric and the scalar field
\begin{eqnarray}
S=\int{\rm d}^4x\sqrt{-g}\Bigl[L_2+L_3+L_4+L_5\Bigr] \label{action1}\,,
\end{eqnarray}
where
\begin{align}
\label{eq:action}
L_2&=K(\phi, X),\\
L_3&=-G_3(\phi, X)\Box\phi,\\
L_4&=G_4(\phi, X)R+G_{4X}\left[(\Box\phi)^2-(\nabla_\mu\nabla_\nu\phi)^2\right],\\
L_5&=G_5(\phi, X)G_{\mu\nu}\nabla^\mu\nabla^\nu\phi-\frac{1}{6}G_{5X}\bigl[(\Box\phi)^3\nonumber\\
&-3\Box\phi(\nabla_\mu\nabla_\nu\phi)^2+2(\nabla_\mu\nabla_\nu\phi)^3\bigr],
\end{align}
where $K$ and $G_i$ are arbitrary functions of $\phi$ and $X\equiv-(\partial\phi)^2/2$, and $G_{iX}\equiv\partial_X G_i$

\section{Background}%%%%%%%%%%%%%%%%%%%%%%%%%%%%%%%%%%%%%%%%%

We consider a static spherically symmetric spacetime for which the metric can be written in the following form
\begin{align}
{\rm d}\bar s^2&=\bar g_{\mu\nu}{\rm d} x^\mu{\rm d} x^\nu\,,\\
\label{eq:metricB}
&=-A(r){\rm d} t^2+\frac{{\rm d} r^2}{B(r)}+C(r)\left({\rm d}\theta^2+\sin^2\theta\,{\rm d} \varphi^2\right).
\end{align}
We could choose the most generic case where the energy-momentum tensor of the scalar field shares the symmetries of the geometry and not the scalar field but, as a matter of simplicity, we will consider that the scalar field inherits the spacetime symmetries, which implies $\phi=\phi(r)$. We will not substitute in our calculations $C=r^2$ because various solutions in the literature can not be written in coordinate system where $C=r^2$, like the Fisher solution for a massless quintessence scalar field \cite{Fisher:1948yn}.

Replacing the metric (\ref{eq:metricB}) in the action (\ref{eq:action}) and varying the action with respect to the functions A, B and C, we get three equations of motion which we will denote as ${\cal E}_A=0$, ${\cal E}_B=0$ and ${\cal E}_C=0$ respectively. Their complete expressions are given in the Appendix \ref{appenB}.

\section{Perturbations}%%%%%%%%%%%%%%%%%%%%%%%%%%%%%%%%%%%%%%%%%

Given the background equations of motion, we can now derive the equations of perturbations. These perturbations decompose into two types which separates at linear order, and therefore can be studied separately. One type of perturbations induce a rotation of the black hole while the second type impart no such rotation. They are called axial (or odd) and polar (or even) perturbations respectively. The metric can be decomposed as $g_{\mu \nu}=\bar g_{\mu \nu}+h_{\mu \nu}$, where $h_{\mu \nu}$ represents the infinitesimal perturbations to the static spherically symmetric background. Each perturbation component can be decomposed into spherical harmonics. Notice that odd-parity modes get a factor $(-1)^{l+1}$ under parity transformation $(\theta,\phi)\rightarrow (\pi-\theta,\pi+\phi)$ while even-parity modes acquire a factor $(-1)^l$. In this paper, we will focus on the odd-parity (axial) modes. We have in this case
\begin{align}
h_{tt}&=0\,,\\
h_{ti}&=\sum_{l,m}h_{0,(l,m)}(t,r)\Bigl[0,\frac{1}{\sin \theta}\sum_{l,m}\partial_\phi Y_l^m,-\sin\theta\sum_{l,m}\partial_\theta Y_l^m\Bigr]\,,\\
h_{ij}&=\sum_{l,m}\Bigl[h_{1,(l,m)}(t,r)(e_1)_{ij}+h_{2,(l,m)}(t,r)(e_2)_{ij}\Bigr]\,,
\end{align}
where the tensor spherical harmonics $(e_{1,2})_{ij}$ are given by
\begin{align}
(e_1)_{ij}=  
\begin{pmatrix}
    0 & \frac{1}{\sin\theta}\partial_\phi Y_l^m & - \sin\theta\partial_\theta Y_l^m \\
    \frac{1}{\sin\theta}\partial_\phi Y_l^m & 0 & 0\\
    - \sin\theta\partial_\theta Y_l^m & 0 & 0
  \end{pmatrix}
\end{align}
and
\begin{multline}
(e_2)_{ij}= \left(
  \begin{matrix}
    0 & 0\\ 
    0 & -\frac{1}{\sin\theta}\Bigl[\partial_{\theta\phi}^2-\cot\theta\partial_\phi\Bigr]Y_l^m \\ 
    0 & \frac{1}{2}\Bigl[-\frac{1}{\sin^2\theta}\partial_\phi^2+\cos\theta\partial_\theta+\sin\theta\partial_\theta^2\Bigr]Y_l^m \\ 
  \end{matrix}\right.                
\\
  \left.
  \begin{matrix}
    0 \\ 
    \frac{1}{2}\Bigl[-\frac{1}{\sin^2\theta}\partial_\phi^2+\cos\theta\partial_\theta+\sin\theta\partial_\theta^2\Bigr]Y_l^m\\
    \Bigl[\sin\theta\partial_{\theta\phi}^2-\cos\theta\partial_\phi\Bigr]Y_l^m
  \end{matrix}\right)\,.
\end{multline}

This can be written in a more compact way by introducing $E_{ab}\equiv \sqrt{\det\gamma}~\epsilon_{ab}$ where $a, b$ $\in \{\theta,\phi\}$, $\gamma_{ab}$ being the metric on the 2-sphere, and $\epsilon_{ab}$ the totally anti-symmetric symbol with $\epsilon_{\theta\phi}=1$. We have then
\begin{align}
h_{ta}&=\sum_{l,m}h_{0,(lm)}(t,r)E_{ab}\partial^bY_l^m(\theta,\phi)\,,\\
h_{ra}&=\sum_{l,m}h_{1,(lm)}(t,r)E_{ab}\partial^bY_l^m(\theta,\phi)\,,\\
h_{ab}&=\frac{1}{2}\sum_{l,m}h_{2,(lm)}\Bigl[E_a^{~c}\nabla_{cb}Y_l^m(\theta,\phi)+E_b^{~c}\nabla_{ca}Y_l^m(\theta,\phi)\Bigr]\,.
\end{align}
Not all perturbations are physical, we can fix some perturbations to zero because of the invariance under arbitrary differentiable coordinate transformations. We consider an infinitesimal coordinate transformation in terms of a small and arbitrary displacement four-vector $\xi^\mu$, $x^\mu\rightarrow x^\mu+\xi^\mu$. Because any scalar, can be decomposed directly into a sum of spherical harmonics and any vector can be decomposed into a divergence part and a divergence-free part. Therefore we have
\begin{align}
\xi_t &=\sum_{l,m}T_{l,m}(t,r)Y_l^m(\theta,\phi)\,,\\
\xi_r &=\sum_{l,m}R_{l,m}(t,r)Y_l^m(\theta,\phi)\,,\\
\xi_a &=\sum_{l,m}\Bigl[\Theta_{l,m}(t,r)\partial_a Y_l^m(\theta,\phi)+\Xi_{l,m}(t,r)E_a^{~b}\partial_b Y_l^m(\theta,\phi)\Bigr]\,.
\end{align}
Notice that $\Theta_{l,m}(t,r)\partial_a Y_l^m(\theta,\phi)$ is the divergence part of the 2-vector $\xi_a$ and $\Xi_{l,m}(t,r)E_a^{~b}\partial_b Y_l^m(\theta,\phi)$, the divergence-free part. Only the latter contributes to the odd-parity perturbations. Therefore we consider an infinitesimal displacement of the following form
\begin{align}
\xi_t &=0\,,\\
\xi_r &=0\,,\\
\xi_a &=\sum_{l,m}\Xi_{l,m}(t,r)E_a^{~b}\partial_b Y_l^m(\theta,\phi)\,.
\end{align}
Under this transformation, the metric perturbations transform as
\begin{align}
h_{0,(l,m)} &\rightarrow h_{0,(l,m)} +\dot\Xi_{l,m}(t,r)\,,\\
h_{1,(l,m)} &\rightarrow h_{1,(l,m)} +r^2\Bigl(\frac{\Xi_{l,m}(t,r)}{r^2}\Bigr)'\,,\\
h_{2,(l,m)} &\rightarrow h_{2,(l,m)} +2 \Xi_{l,m}(t,r)\,.
\end{align}
We see therefore that the only way to fix completely the gauge is to consider $h_2=0$, any other choice, will fix partially the gauge which is usually refereed to as the ``Regge-Wheeler'' gauge \cite{Regge:1957td}. The additional advantage of using this gauge is that the equations of motion are independent of $m$, which can therefore be set to zero. Therefore the spherical harmonics can be replaced by Legendre polynomials when fixing $m=0$
\begin{align}
Y_l^m=\sqrt{\frac{(2l+1)(l-m)!}{4\pi(l+m)!}}P_l^me^{im\phi} \xrightarrow[m=0]{} \sqrt{\frac{(2l+1)}{4\pi}}P_l(\cos \theta)\,,
\end{align}
where $P_l^m$ are the associated Legendre polynomials and $P_l$, the Legendre polynomials. Finally, the equations of motion for different $l$ decouple and we can therefore pick up a single value for $l$. In summary, for this particular gauge choice, the odd-parity metric perturbations assume the simple form
\begin{equation}
h_{\mu \nu}=  -
\begin{pmatrix}
    0 & 0 & 0 & h_0\\
    0 & 0 & 0 & h_1\\
    0 & 0 & 0 & 0\\
    h_0 & h_1 & 0 & 0\\
  \end{pmatrix}  \sqrt{\frac{(2l+1)}{4\pi}}\sin\theta P'_l(\cos \theta)\,,
\end{equation}
where we simplified the notation $h_0=h_{0,(l,m)}$ and $h_1=h_{1,(l,m)}$ since $(l,m)$ are fixed. 

In addition to the metric perturbations, we also need to perturb the scalar field, $\phi=\bar\phi(r)+\delta\phi(t,r,\theta,\phi)$. Depending on the context, we assume that there will be no confusion between the scalar field and the azimuthal angle. This perturbation can also be expanded into spherical harmonics
\begin{align}
\delta\phi=\sum_{l,m}\Phi(t,r)Y_l^m(\theta,\phi)\,.
\end{align}
Scalar field perturbation will contribute only to even-parity modes and therefore we will consider $\delta\phi=0$.

\subsection{Second-order action}%%%%%%%%%%%%%%%%%%%%%

To obtain a linear equation for $h_0$ and $h_1$, we expand the action (\ref{action1}) to second order in the perturbations. We derive the following action:
\begin{eqnarray}
S^{(2)}=\int {\rm d} t\,{\rm d} r\,{\cal L}^{(2)}\,,\label{action-init}
\end{eqnarray}
where we performed an integration over the angles $(\theta,\phi)$ by using the standard relations of the Legendre Polynomials and multiple integration by parts. The second-order Lagrangian ${\cal L}^{(2)}$ is of the form
\begin{equation}
{\cal L}^{(2)}=a_1 h_0^2+a_2 h_1^2+a_3 \Bigl[ {\dot h_1}^2+h_0'^2-2 {\dot h_1} h_0'+2\frac{C'}{C} {\dot h_1} h_0 \Bigr]\,, \label{lag-odd1}
\end{equation}
where a dot represents differentiation with respect to $t$ and $'$ with respect to $r$. The coefficients $a_1, a_2$ and $a_3$ are given by
\begin{align}
a_1&=\frac{\ell (\ell+1)}{4 C}
\Bigl[\frac{{\rm d}}{{\rm d} r}\left(
C'\sqrt{\frac{B}{A}}{\cal H}\right)
+\frac{(\ell-1) (\ell+2)}{\sqrt{AB}}{\cal F}\nonumber\\
&+\frac{2C}{\sqrt{AB}}  {\cal E}_A
\Bigr]\,,
\\
a_2&=- \frac{\ell (\ell+1)}{2} \sqrt{AB} \left[ \frac{(\ell-1) (\ell+2)}{2 C}{\cal G}+{\cal E}_B \right]\,,
\\
a_3&=
\frac{\ell (\ell+1)}{4} \sqrt{\frac{B}{A}} {\cal H}\,.
\end{align}
On-shell, ${\cal E}_A={\cal E}_B=0$ and hence the above expressions simplify a bit
\begin{align}
{\cal F}&=2\left(
G_4+\frac{1}{2}B\phi'X'G_{5X}-XG_{5\phi}
\right)\,, \label{def-f}
\\
{\cal G}&=2\left[ G_4-2XG_{4X}
+X\left(\frac{A'}{2A}B\phi'G_{5X}+G_{5\phi}\right)
\right]\,, \label{def-g}
\\
{\cal H}&=2\left[G_4-2X G_{4X}
+X\left(\frac{  C' }{2 C} B \phi' G_{5X}+ G_{5\phi}\right)\right]\,. \label{def-h}
\end{align}
and therefore we recover the same results as \cite{Kobayashi:2012kh}.

The Lagrangian (\ref{lag-odd1}) can be simplified by noticing that the ${\dot h_0}$ is absent and therefore it is an auxiliary field. But it is complicated to integrate it out using its equation of motion because of the presence of the term $h_0'^2$ in the Lagrangian. Therefore, we introduce an additional field \cite{DeFelice:2011ka} by rewriting the Lagrangian in the following form
\begin{equation}
{\cal L}^{(2)}=\Bigl[ a_1-\frac{(C' a_3)'}{C} \Bigr] h_0^2
+a_2 h_1^2+a_3
\Bigl[ \dot h_1-h_0'+\frac{C'}{C}h_0 \Bigr]^2\,. \label{eq:Lagr}
\end{equation}
We now introduce an auxiliary field q and define the following Lagrangian
\begin{align}
\label{eq:Lagq}
L&=\Bigl[ a_1-\frac{(C' a_3)'}{C} \Bigr] h_0^2
+a_2 h_1^2+a_3
\Bigl[ 2q \left( {\dot h_1}- h_0'+\frac{C'}{C} h_0 \right) \nonumber\\
& -q^2\Bigr]\,.
\end{align}
It is easy to see that both the Lagrangians give the same equations of motion, in fact, performing a variation with respect to $q$ gives
\begin{align}
q={\dot h_1}- h_0'+\frac{C'}{C} h_0\,,
\end{align}
which when replaced in (\ref{eq:Lagq}) gives (\ref{eq:Lagr}). Therefore, varying the Lagrangian (\ref{eq:Lagq}) with respect to $h_0$ and $h_1$, leads to
\begin{equation}
h_0=-\frac{\left( C a_3 q \right)'}{C a_1-\left( C' a_3 \right)'}\,,~~~h_1=\frac{a_3}{a_2} {\dot q}\,,\label{hhq}
\end{equation}
respectively. Plugging this result back to (\ref{eq:Lagq}), we obtain a quadratic Lagrangian solely in terms of the master variable $q$, which constitutes the only degree of freedom$-$the sole propagating odd mode. Any other function, such as $h_0$ and $h_1$, can be determined once $q$ is known using the relations (\ref{hhq}). Solving few integration by parts, we arrive to a canonical form of the action
\begin{align}
\label{eq:final2}
S^{(2)}=\int {\rm d} t\,{\rm d} r\,\Bigl[\alpha \dot q^2+\beta q'^2+\gamma q^2 \Bigr]\,,
\end{align}
where
\begin{align}
\alpha&=\frac{l(l+1)}{4(l-1)(l+2)}\sqrt{\frac{B}{A}}\frac{C}{A}\frac{{\cal H}^2}{{\cal G}}\,,\\
\beta&=-\frac{l(l+1)}{4(l-1)(l+2)}\sqrt{\frac{B}{A}}BC\frac{{\cal H}^2}{{\cal F}}\,,
\end{align}
while $\gamma$ has a longer expression and will not be necessary at this level. In order to avoid a ghost, we need to impose $\alpha>0$, which means ${\cal G}>0$ and the condition to avoid Laplacian instabilities is $\beta<0$ or ${\cal F}>0$, which corresponds to real sound speed.

To arrive at the final result, we redefine the variable $q$ as
\begin{equation}
q=\sqrt{\frac{A {\cal F}}{ BC {\cal H}^2}} Q\,,
\label{Qq}
\end{equation}
and introduce the tortoise coordinate, ${\rm d}r=\sqrt{AB}{\rm d}r^{*}$. The action (\ref{eq:final2}) finally takes the form
\begin{align}
\label{eq:final}
S^{(2)}=\frac{l(l+1)}{4(l-1)(l+2)}\int {\rm d} t\,{\rm d} r^{*}\,\Bigl[\frac{{\cal F}}{{\cal G}} \dot Q^2-\Bigl(\frac{dQ}{dr^{*}}\Bigr)^2-V(r) Q^2 \Bigr]\,,
\end{align}
where the potential is defined as
\begin{align}
\label{waveeq}
V&=l(l+1)\frac{A}{C}\frac{{\cal F}}{{\cal H}}
-\frac{C^2}{4C'}\Bigl(\frac{ABC'^2}{C^3}\Bigr)'
-\frac{C^2{\cal F}^2}{4{\cal F}'}\Bigl(\frac{AB{\cal F}'^2}{C^2{\cal F}^3}\Bigr)'\nonumber\\
&-\frac{2A{\cal F}}{C{\cal H}}\,.
\end{align}
Even if the final result is structurally similar to \cite{Kobayashi:2012kh}, we differ in the form of the potential.

Even if the action is not defined for $l=0$ and $l=1$, these modes are not dynamical. For instance, $l=0$ is spherically symmetric and therefore obey the Birkhoff theorem, while $l=1$ corresponds to an infinitesimal shift of the position of the black hole \cite{Regge:1957td}.

Since $\ell (\ell+1)$ corresponds to the two-dimensional Laplacian in the real space, the first term in Eq. (\ref{waveeq}) represents wave propagation along the angular direction. Therefore, we also impose ${\cal H}>0$, which gives positive squared propagation speeds along the angular direction. In summary, the avoidance of ghost and Laplacian instability demands the conditions 
\begin{align}
\label{eq:stab}
{\cal F}>0,~~{\cal G}>0,~~{\cal H}>0\,.
\end{align}

\section{Stability}
From the variation of the action (\ref{eq:final}), we find
\begin{align}
\label{eq:master}
-\frac{\partial^2 Q}{\partial t^2}+\frac{\cal G}{\cal F}\frac{\partial^2 Q}{\partial r^{*2}}-\frac{\cal G}{\cal F} V Q=0\,.
\end{align}
Therefore, we see that we can define the speed of fluctuations propagating radially as
\begin{align}
c_r^2=\frac{\cal G}{\cal F}\,.
\end{align}
Also, the term $\frac{\cal G}{\cal F} VQ$, includes the angular part, which from (\ref{waveeq}), is
\begin{align}
l(l+1)\frac{A}{C}\frac{\cal G}{\cal H}\,.
\end{align}
This comes from the angular part of the Laplacian, from which we can define the speed of angular excitations
\begin{align}
c_\Omega^2=\frac{\cal G}{\cal H}\,.
\end{align}
We therefore ensure the positivity of the speed squared of propagation of the perturbations due to the conditions given in (\ref{eq:stab}). Having the equation of perturbations, we now turn to their stability. Once a black hole is perturbed, it responds to perturbations by emitting gravitational waves (GWs). This signal can be divided into three stages
\begin{enumerate}
\item An early response which depends strongly on the initial conditions.
\item An exponentially decaying phase, known as the ringdown.
\item A late tail.
\end{enumerate}
The distortion of the solution reduces to the spherical solution after the emission of GWs during these stages. The linear theory of perturbations studied in this paper tackles the last two stages of this process, during which the waveform is typically identified with a quasinormal frequency $\omega$ and therefore this exponential damping of the perturbations is called quasi-normal ringing. Because the ringdown signal is given by a superposition of quasinormal modes (QNMs), we can write
\begin{align}
\label{Fourier}
Q(t,r)=\sum_n e^{-\iu \omega_n t}\psi_n(r)\,.
\end{align}
Replacing in (\ref{eq:master}), we have
\begin{align}
\label{eq:master2}
\Bigl(-\frac{d^2}{dr^{*2}}+V\Bigr)\psi_n=\frac{\omega_n^2}{c_r^2}\psi_n\,.
\end{align}
Any frequency $\omega_n$ solves this equation but only a discrete number solves it with additional boundary conditions dictated by the physical problem. The boundary conditions defining QNMs are purely ingoing at the horizon and outgoing at infinity for asymptotically flat or de Sitter (dS) spacetimes and should be null at infinity for AdS spacetime. These conditions are physically motivated, e.g., ingoing at the horizon means entering into the black hole and we should not have any perturbations outgoing at the horizon. Equivalently, we should not have a wave coming from infinity. These frequencies are usually complex and we can see from (\ref{Fourier}) that the modes are not growing in time when $\operatorname{Im}{(\omega_n)}<0$. Therefore the stability of the black hole under linear perturbations dictate that all QNMs should fulfill the condition $\operatorname{Im}{(\omega_n)}<0$. 

Multiplying eq. (\ref{eq:master2}) by the complex conjugate of $\psi_n$ and integrating over the tortoise coordinate, we obtain after an integration by parts
\begin{align}
\label{eq:master3}
-\bar \psi_n \frac{d\psi_n}{dr^*}\Bigl |_{-\infty}^{+\infty}+\int_\R {\rm d}r^* \Bigl[\Bigl |\frac{d\psi_n}{dr^*}\Bigr |^2+V |\psi_n |^2\Bigr]=\omega_n^2 A^2\,,
\end{align}
where $\bar \psi_n$ is the complex conjugate of $\psi_n$ and $A^2=\int_\R {\rm d}r^* \frac{ |\psi_n |^2}{c_r^2}>0$. In the first situation, where the black hole is asymptotically flat or dS, the choice of the boundary conditions are ingoing near the horizon $(r^*=-\infty)$ and outgoing at infinity or at the cosmological horizon $(r^*=+\infty)$. Therefore,
\begin{align}
\psi_n(r^*)&\propto e^{-\iu \omega_n r^*}\,,~~\text{when  }r^*=-\infty~~\text{and}\nonumber\\
\psi_n(r^*)&\propto e^{+\iu \omega_n r^*}\,,~~\text{when  }r^*=+\infty\,.
\end{align}
This implies that there is no signal which comes from the black hole or from any source at infinity. The conditions for asymptotically AdS are the same at the horizon but the Dirichlet one at infinity, given by
\begin{align}
\psi_n(r^*)&\propto e^{-\iu \omega_n r^*}\,,~~\text{when  }r^*=-\infty~~\text{and}\nonumber\\
\psi_n(r^*)&=0\,,~~\text{when  }r^*=0
\end{align}
where $r^*=0$ corresponds to $r=\infty$.

We therefore have
\begin{align}
-\bar \psi_n \frac{d\psi_n}{dr^*}\Bigl |_{-\infty}^{+\infty}=-\iu \omega_n B^2\,,
\end{align}
where $B^2=|\psi_n(+\infty)|^2+|\psi_n(-\infty)|^2$, for an asymptotically flat or dS black hole and $B^2=|\psi_n(-\infty)|^2$ for an asymptotically AdS spacetime. Hence, $B^2$ is always a real positive number. Therefore, the eq. (\ref{eq:master3}) becomes
\begin{align}
\label{eq:AB}
\int_\R {\rm d}r^* \Bigl[\Bigl |\frac{d\psi_n}{dr^*}\Bigr |^2+V |\psi_n |^2\Bigr]=\omega_n^2 A^2+\iu \omega_n B^2\,.
\end{align}
The imaginary part of this equation reads
\begin{align}
\operatorname{Re}{(\omega_n)} \Bigl(2 \operatorname{Im}{(\omega_n)}A^2+ B^2\Bigr)=0\,.
\end{align}
Therefore, if the black hole is unstable (i.e., $\operatorname{Im}{(\omega_n)}>0$) then $\operatorname{Re}{(\omega_n)}=0$. The unstable modes do not oscillate. So, the unstable modes are equivalent to the condition $\omega_n^2<0$ (i.e., purely imaginary modes), which from (\ref{eq:master2}) concludes that the stability of the spacetime is related to the positivity of the operator $-\frac{d^2}{dr^{*2}}+V$.

We now turn to the real part of (\ref{eq:AB}) which reads
\begin{align}
\label{eq:AB2}
\int_\R {\rm d}r^* \Bigl[\Bigl |\frac{d\psi_n}{dr^*}\Bigr |^2+V |\psi_n |^2\Bigr]&=(\operatorname{Re}{(\omega_n)}^2-\operatorname{Im}{(\omega_n)}^2)A^2\nonumber\\
&-\operatorname{Im}{(\omega_n)}B^2\,.
\end{align}
If the mode is unstable, we have $\operatorname{Re}{(\omega_n)}=0$ and $\operatorname{Im}{(\omega_n)}>0$, which implies the right-hand side of eq. (\ref{eq:AB2}) being negative, and therefore this condition can be fulfilled only if the potential on the left-hand side of eq. (\ref{eq:AB2}) is negative. Of course, a negative potential can yet give a positive integral. On the contrary, if the potential is positive, we necessarily have stability. In the case of a negative potential, the ``S-deformation'' approach is useful. For that, we introduce a smooth function and a new ``derivative'', $D=\frac{d}{dr^*}+S$. We get, after integration by parts
\begin{align}
\label{eq:fin}
\int_\R {\rm d}r^* \Bigl[\Bigl |\frac{d\psi_n}{dr^*}\Bigr |^2+V |\psi_n |^2\Bigr]&=\int_\R {\rm d}r^* \Bigl[ |D\psi_n |^2+W |\psi_n |^2\Bigr]\nonumber\\
&-S|\psi_n |^2\Bigl |_{r^{*}=-\infty}^{r^{*}=+\infty}\,,
\end{align}
where the new potential $W$ is defined as
\begin{align}
W=V+\frac{dS}{dr^*}-S^2\,.
\end{align}
We therefore see that if we can find a function $S$ such as $S(r^*=+\infty)\le 0$, $S(r^*=-\infty)\ge 0$ and $W\ge0$, we will have the left-hand side of eq. (\ref{eq:AB2}) positive and hence stability of the solution. 

Considering 
\begin{align}
S=\frac{1}{2}\frac{d \ln (C\cal F)}{dr^*}=\frac{\sqrt{AB}}{2}\Bigl(\frac{C'}{C}+\frac{\cal F'}{\cal F}\Bigr)\,,
\end{align}
we find that
\begin{align}
W=(l+2)(l-1)\frac{A}{C}\frac{\cal F}{\cal H}\,.
\end{align}
If ${\cal F}/{\cal H}<0$, the solution is unstable. In fact, eq. (\ref{eq:fin}) is necessarily negative for sufficiently large $l$ (because of $W$), while if ${\cal F}/{\cal H}>0$ and $S(r^*=+\infty)\le 0$, $S(r^*=-\infty)\ge 0$, all the terms of eq. (\ref{eq:fin}) are positive for any $l$ which ensures the stability.

In conclusion, Horndeski black holes are linearly stable under odd perturbations if and only if ${\cal F}>0$, ${\cal G}>0$ and ${\cal H}>0$. Also we need the conditions $S(r^*=+\infty)\le 0$, $S(r^*=-\infty)\ge 0$ where
\begin{align}
\label{eq:S-def}
S=\frac{\sqrt{AB}}{2}\Bigl(\frac{C'}{C}+\frac{\cal F'}{\cal F}\Bigr)\,.
\end{align}

\section{Application to particular models}%%%%%%%%%%%%%%%%%%%%%%%%%%%%%%%%%%%%%%%%%

\subsection{General Relativity}%%%%%%%%%%%%%%%%%%%%%

We first apply the results to the most standard example to check the consistency of our calculations. In general relativity, the action can be written in the form (\ref{eq:action}) with

\begin{equation}
K=0,~G_3=0,~G_4=\frac{1}{2},~G_5=0\,.
\end{equation}
In this case, we have 
\begin{equation}
{\cal F}={\cal G}={\cal H}=1\,,
\end{equation}
which are all positive. The stability is therefore proved iff the S-deformation technique used is licit, which means if the conditions $S(r^*=+\infty)\le 0$ and $S(r^*=-\infty)\ge 0$ is true for $S$ defined in (\ref{eq:S-def}). For GR, it is easy to calculate that 
\begin{align}
S=\frac{A(r)}{r}\,,\quad\text{where}  ~~A=1-\frac{2M}{r}\,.
\end{align}
We consider the Schwarzschild solution
\begin{align}
A=B=1-\frac{2M}{r}\,,~~C=r^2\,,
\end{align}
and find that  $S(r^*=+\infty)=S(r^*=-\infty)=0$ where $r^*$ is the tortoise coordinate, which allows us to conclude that Schwarzschild spacetime is stable under odd perturbations. For completeness, we derive the potential $V(r)$ from eq. (\ref{waveeq})
\begin{align}
V(r)=\Bigl(1-\frac{2M}{r}\Bigr)\Bigl[\frac{l(l+1)}{r^2}-\frac{6M}{r^3}\Bigr]\,,
\end{align}
which is the Regge-Wheeler potential \cite{Regge:1957td}.

We conclude that the Schwarzschild spacetime does not exhibit either ghost or gradient instability, it is stable and the propagation speeds along both radial and angular directions are equal to the speed of light, $c_r^2=c_\theta^2=1$.

\subsection{General Relativity conformally coupled to a scalar filed}%%%%%%%%%%%%%%%%%%%%%

Considering a scalar field with non-negative potential in asymptotically flat spacetimes, it is well known that static black holes do not have scalar hair \cite{Bekenstein:1971hc} (see \cite{Herdeiro:2015waa} for a recent review). In fact, if the scalar field is not zero, then the equations show that it will diverge at the horizon, implying a divergence of the curvature too. Therefore to obtain a regular black hole solution, one has to consider the trivial solution where the scalar field vanishes and hence the theory reduces to Einstein gravity in vacuum, imposing that the solution is Schwarzschild, or more generically, Kerr-Newman family. We therefore show the non-existence of hairy black holes. But in 1972, a conformally coupled scalar field to GR giving rise to solution asymptotically flat and different form Schwarzschild was derived. The scalar field is still singular on the horizon but without a divergence of the curvature tensor. The action is defined as
\begin{align}
S=\int{\rm d}^4x\sqrt{-g}\Bigl[\frac{R}{2}-\frac{1}{2}(\partial\phi)^2-\frac{R}{12}\phi^2\Bigr]\,,
\end{align}
where the solution is 
\begin{align}
ds^2 &=-\Bigl(1-\frac{M}{r}\Bigr)^2 dt^2+\frac{dr^2}{\Bigl(1-\frac{M}{r}\Bigr)^2}+r^2d\Omega^2\,,\\
\phi &=\frac{\sqrt{6}M}{r-M}\,.
\end{align}
This is known as the Bocharova-Bronnikov-Melnikov-Bekenstein (BBMB) solution \cite{Bocharova:1970skc,Bekenstein:1974sf}. Even if the solution is not Schwarzschild, it is the extreme Reissner-Nordstr\"{o}m solution, there are no other parameters to describe the solution except the mass. Therefore it is not a hairy black hole solution (or primary hair) but secondary hair solution \cite{Herdeiro:2015waa}.

The stability of this solution is unclear since both situations have been claimed$-$stable in \cite{McFadden:2004ni} and unstable in \cite{Bronnikov:1978mx}. Unfortunately, the stability has been studied only for radial (monopole) perturbations in case of even-parity perturbations. We, therefore, extend the analysis and study completely the odd-parity perturbations. In our notations, for this case, we have 
\begin{equation}
K=X,~G_3=0,~G_4=\frac{1}{2}\Bigl(1-\frac{\phi^2}{6}\Bigr),~G_5=0\,.
\end{equation}
Therefore, we obtain
\begin{equation}
{\cal F}={\cal G}={\cal H}=1-\frac{\phi^2}{6}=\frac{r(r-2M)}{(r-M)^2}\,.
\end{equation}
It is therefore easy to check that the S-deformation is licit, $S(r^*=+\infty)=S(r^*=-\infty)=0$. So, we can directly use our conditions of stability which shows that outside the horizon, in the range $M<r<2M$, we have ${\cal F},~{\cal G}$ and ${\cal H}$ negative. 

In conclusion the BBMB solution is unstable under odd gravitational perturbations. 

\subsection{Extended BBMB solution}%%%%%%%%%%%%%%%%%%%%%

In \cite{Martinez:2002ru}, the authors extended the solution by including a cosmological constant (MTZ solution)
\begin{align}
ds^2 &=-\Bigl[\Bigl(1-\frac{M}{r}\Bigr)^2-\frac{\Lambda}{3}r^2\Bigr] dt^2+\frac{dr^2}{\Bigl(1-\frac{M}{r}\Bigr)^2-\frac{\Lambda}{3}r^2}\nonumber\\
&\qquad\qquad\qquad\qquad\qquad\qquad\qquad\qquad+r^2d\Omega^2\,,\\
\phi &=\frac{\sqrt{6}M}{r-M}\,.
\end{align}
For that, they had to add in the action, a quartic interaction term along with a cosmological constant
\begin{align}
S=\int{\rm d}^4x\sqrt{-g}\Bigl[\frac{R}{2}-\frac{1}{2}(\partial\phi)^2-\frac{R}{12}\phi^2-\Lambda+\frac{\Lambda}{36}\phi^4\Bigr]\,.
\end{align}
This solution is studied between the event horizon $r_\text{min}$ and the cosmological horizon $r_\text{max}$, given by
\begin{align}
r_\text{min} &=\frac{l}{2}\Bigl(1-\sqrt{1-\frac{4M}{l}}\Bigr)\,,\\
r_\text{max} &=\frac{l}{2}\Bigl(1+\sqrt{1-\frac{4M}{l}}\Bigr)\,,
\end{align}
where $l=\sqrt{3/\Lambda}>0$ and hence, the existence of the horizon imposes the condition $0<4M<l$. For this class of theories, we find that the S-deformation is licit, $S(r^*=+\infty)=S(r^*=-\infty)=0$, and the stability condition is given by the positivity of
\begin{equation}
{\cal F}={\cal G}={\cal H}=1-\frac{\phi^2}{6}=\frac{r(2-2M)}{(r-M)^2}\,.
\end{equation}
It is again easy to check that the odd perturbations are unstable because ${\cal F},~{\cal G}$ and ${\cal H}$ are negative in the range $r_\text{min}<r<2M$. In fact in the range of existence of the black hole solution, $0<4M<l$, we always have $r_\text{min}<2M$.

\subsection{No scalar-hair theorem}%%%%%%%%%%%%%%%%%%%%%

Various no-hair theorems were derived in the literature, where the only static spherical regular black hole is proved to be Schwarzschild. For that, the authors show that the scalar field should be trivial (i.e., constant or zero). We can, for example, cite the no hair theorem for massless scalar field \cite{chase}, minimally coupled scalar field with arbitrary positive potential \cite{Heusler:1992ss,Sudarsky:1995zg}, nonminimally coupled scalar field \cite{Mayo:1996mv,Bekenstein:1996pn}, in Brans-Dicke theory of gravity \cite{Faraoni:2013iea} or more recently, for galileons \cite{Hui:2012qt}. Considering that we have for these models Schwarzschild as a solution with a constant or vanishing scalar field $(\phi_0)$, we conclude that 
\begin{equation}
{\cal F}={\cal G}={\cal H}=2G_4(\phi_0,0)\,,
\end{equation}
and therefore, Schwarzschild solution is stable in these models if $G_4(\phi_0,0)>0$. Of course, for that we need to consider the S-deformation technique is licit. We will see in the next example, a model for which we can't use this approach.

\subsection{Non-minimal derivative coupling}

\begin{figure}[H]
\begin{center}
\includegraphics[scale=.8]{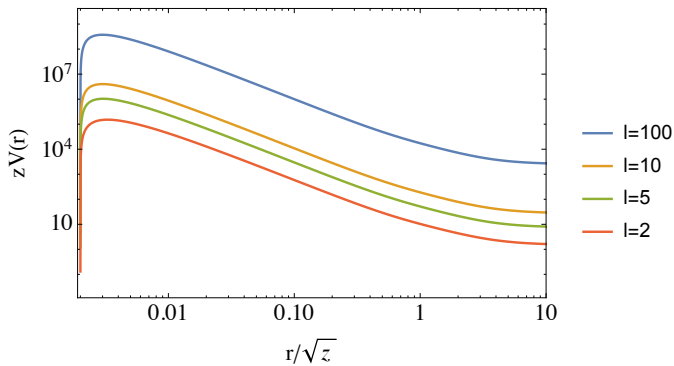}
\end{center}
\caption{Potential for the black hole solution (\ref{eq:Rinaldi1},\ref{eq:Rinaldi2}) from the horizon where the potential is null to its asymptotic value $(9l(l+1)-1)/36 z$.}
\label{fig:rinaldi}
\end{figure}

Among the possible models of Horndeski, the coupling between the derivative of the scalar field and the Einstein tensor has also been studied. In this case the action is
\begin{align}
S=\int{\rm d}^4x\sqrt{-g}\Bigl[\frac{R}{2}+\frac{1}{2}(\partial\phi)^2+\frac{z}{2}\phi G_{\mu\nu}\nabla^\mu\nabla^\nu\phi\Bigr]\,,
\end{align}
and the black hole solution was found in \cite{Rinaldi:2012vy}, which is
\begin{align}
\label{eq:Rinaldi1}
A&=\frac{3}{4}-\frac{2M}{r}+\frac{r^2}{12z}+\frac{\sqrt{z}}{4r}\arctan\Bigl(\frac{r}{\sqrt{z}}\Bigr)\,,\\
\label{eq:Rinaldi2}
B&=\frac{4(r^2+z)^2A}{(r^2+2z)^2}\,,\\
\phi'^2&=\frac{r^2(r^2+2z)^2}{4z(r^2+z)^3A}\,.
\end{align}
Notice the (+) sign of the kinetic term in the action which we adopt in order to have a real scalar field while it is defined as complex in \cite{Rinaldi:2012vy}. Also, in the limit $z\rightarrow \infty$, the scalar field becomes constant and the metric becomes Schwarzschild. Therefore, the solution constitutes a smooth deformation of the Schwarzschild spacetime. We have in our notations
\begin{align}
K=-X,~G_3=0,~G_4=\frac{1}{2},~G_5=\frac{z}{2}\phi\,,
\end{align}
which gives
\begin{align}
{\cal F}&=\frac{3r^2+2z}{2(r^2+z)}\,,\\ 
{\cal G}&={\cal H}=\frac{r^2+2z}{2(r^2+z)}\,.
\end{align}
We might conclude that the solution is stable because ${\cal F}>0$, ${\cal G}>0$ and ${\cal H}>0$ but, in fact, the S-deformation is not licit, since $S(r^*=0)=+\infty$ and therefore it is an example where the formulas derived can't be used. We study the potential $V$ directly from which it is easy to see that $V(r^*=-\infty)=0$ and $V(r^*=0)=\frac{9l(l+1)-1}{36 z}$, and we have checked graphically that the potential is always positive (see Fig.\ref{fig:rinaldi}), which implies the stability of the solution.

\section{Conclusion}%%%%%%%%%%%%%%%%%%%%%%%%%%%%%%%%%%%%%%%%%

In this paper, we have studied odd-parity perturbations around static and spherically symmetric background spacetime in Horndeski gravity. We have derived the conditions of no-ghost and Laplacian instability, as well as the conditions of stability under these perturbations. We found that these conditions are similar and reduce to very simple conditions
\begin{align}
{\cal F}(r)>0,\quad {\cal G}(r)>0,\quad {\cal H}(r)>0\,,\nonumber
\end{align}
where $r$ runs from the event horizon to infinity or cosmological horizon and
\begin{align}
{\cal F}&=2\Bigl[
G_4+\frac{1}{2}B\phi'X'G_{5X}-XG_{5\phi}
\Bigr]\,,\nonumber
\\
{\cal G}&=2\left[ G_4-2XG_{4X}
+X\left(\frac{A'}{2A}B\phi'G_{5X}+G_{5\phi}\right)
\right]\,,\nonumber
\\
{\cal H}&=2\left[G_4-2X G_{4X}
+X\left(\frac{  C' }{2 C} B \phi' G_{5X}+ G_{5\phi}\right)\right]\,.\nonumber
\end{align}
We have applied the results to various solutions encountered in the literature. We found that for some spacetimes, the S-deformation is not possible because of a boundary term which makes the argument impossible. The condition to use the S-deformation and therefore the previous results is $S(r=r_{\text{Horizon}})\geq 0$ and $S(r=+\infty)\leq 0$ where $+\infty$ can be the cosmological horizon in the case of an asymptotically de Sitter solution.
\begin{align}
S=\frac{\sqrt{AB}}{2}\Bigl(\frac{C'}{C}+\frac{{\cal F}'}{{\cal F}}\Bigr)
\end{align}
The condition will often be violated in asymptotically AdS solution, and therefore this S-deformation will not be used for these spacetimes. In these particular cases, we can study directly the potential and show its positivity as we have done in this paper or use an other S-deformation.

\section*{Acknowledgments}

A. Ganguly wants to thank Claude Leon Foundation for financial support. The work of R. Gannouji is partially supported by DII-PUCV No 039.450/2017 and Fondecyt project No 1171384. M. Gonzalez-Espinoza acknowledges support from a PUCV doctoral scholarship.

\appendix

\section{Verification of the paper \cite{Kobayashi:2012kh}}%%%%%%%%%%%%%%%%%%%%%%%%%%%%%%%%%%%%%%%%%
\label{appenA}

In \cite{Kobayashi:2012kh}, the authors performed an analysis of black hole perturbations in the Horndeski model under odd-parity perturbations. In this appendix, we will show that in the simplest case of Schwarzschild solution in GR, their result do not produce the famous Regge and Wheeler potential. 

Considering GR and therefore, ${\cal F}={\cal G}={\cal H}=1$ and $A=B=1-2M/r$, the second order action they found gives
\begin{align}
S\propto \int{\rm d}t{\rm d}r\Bigl[\frac{\dot Q^2}{A^2}-Q'^2-\frac{l(l+1)}{r^2A}Q^2-VQ^2\Bigr]\,,
\end{align}
where
\begin{align}
V(r)=\frac{M(61M-32r)}{4r^2(r-2M)^2}\,.
\end{align}
Transforming the coordinate $r$ to the tortoise coordinate $dr=A dr^*$ and considering the change of variable $Q=\sqrt{A}\xi$, we get after an integration by parts
\begin{align}
S\propto \int{\rm d}t{\rm d}r^*\Bigl[\dot\xi^2-\Bigl(\frac{d\xi}{dr^*}\Bigr)^2-W\xi^2\Bigr]\,,
\end{align}
which gives after variation
\begin{align}
-\frac{\partial^2 \xi}{\partial t^2}+\frac{\partial^2 \xi}{\partial r^{*2}}-W\xi=0\,,
\end{align}
where
\begin{align}
W(r)=A\Bigl[\frac{l(l+1)}{r^2}+\frac{M(41M-24r)}{4r^3(r-2M)}\Bigr]\,,
\end{align}
which is clearly not the Regge-Wheeler potential. 

Finally, it can easily be checked that our result is generalizing previous calculations derived for specific models (see e.g. \cite{Anabalon:2014lea,Cisterna:2015uya}).

\section{Background equations}%%%%%%%%%%%%%%%%%%%%%%%%%%%%%%%%%%%%%%%%%
\label{appenB}
Here we define the quantities that appear in the background equations ${\cal E}_A={\cal E}_B={\cal E}_C=0$
\begin{widetext}
\begin{eqnarray}
{\cal E}_A&:=&K+B\phi'X'G_{3X}-2XG_{3\phi}
+
\frac{2}{C}\left(1+\dfrac{BC'^2}{4C}-BC''-\dfrac{B'C'}{2}\right)  G_4
+
\frac{2BC'}{C}\left( 
\frac{2C''}{C'}-\dfrac{C'}{2C}+\frac{X'}{X}+\frac{B'}{B} \right) X G_{4X}
\nonumber\\&&
+\frac{4BC'}{C}XX'G_{4XX}-B\phi'\left(\frac{2C'}{C}+\frac{X'}{X}\right)
G_{4\phi}
+4X G_{4\phi\phi}
+2B\phi' \left(\frac{2C'}{C}-\frac{X'}{X} \right) X  G_{4\phi X}
\nonumber\\&&
+
\frac{B\phi' }{C}
\left[(1-\dfrac{3BC'^2}{4C})\frac{X'}{X}+\dfrac{BC'}{C}\left(\dfrac{C'^2}{2C}-C''\right)  -\dfrac{B'C'^2}{2C} \right] X G_{5X}
-\frac{B^2 C'^2 \phi'}{2C^2} XX'G_{5XX}
\nonumber\\&&
-
\frac{2}{C}\left[1+B\left(C'' -\frac{C'^2}{4C}\right)+BC'\frac{X'}{X}+\dfrac{B'C'}{2}
\right] X  G_{5\phi}
-\frac{2BC'\phi'}{C} XG_{5\phi\phi}
+\frac{2BC'}{C}\left(\frac{C'}{2C}-\frac{X'}{X} \right)X^2 
G_{5\phi X},
\\
{\cal E}_B&:=&K-2XK_X
+\left(\frac{2C'}{C}+\frac{A'}{A}\right)B \phi' XG_{3X}+2X G_{3\phi}
\nonumber\\&&
+\frac{2}{C}\left(1-\dfrac{BC'^2}{4C}-\dfrac{A'BC'}{2A}\right)G_4
-\frac{4}{C}\left(1-\dfrac{BC'^2}{2C}-\frac{A'BC'}{A}\right)XG_{4X}
+\frac{4BC'}{C}\left(\frac{C'}{2C}+\frac{A'}{A}\right)X^2G_{4XX}
\nonumber\\&&
-\left(\frac{2C'}{C}+ \frac{A'}{A}\right)B\phi'G_{4\phi}
-2\left(\frac{2C'}{C}+ \frac{A'}{A}\right)B\phi'XG_{4\phi X}
+\frac{B\phi'}{C}\left(1-\dfrac{5BC'^2}{4C}\right)\frac{A'}{A}XG_{5X}
\nonumber\\&&
-\frac{A'B^2C'^2 \phi'}{2AC^2}X^2G_{5XX}
+\frac{2}{C}\left(1-\dfrac{3BC'^2}{4C}-\dfrac{3A'BC'}{2A}\right)XG_{5\phi}
-\frac{2BC'}{C}\left(\frac{C'}{2C}+ \frac{A'}{A}\right)X^2G_{5\phi X},
\\
{\cal E}_C&:=&K-2XG_{3\phi}+B\phi' X' G_{3X} -\left[ \sqrt{\frac{B}{A}} \left( \sqrt{\frac{B}{A}} A' \right)'+\sqrt{\frac{B}{C}} \left( \sqrt{\frac{B}{C}} C' \right)'+\frac{A'BC'}{2AC} \right]G_4
\nonumber\\&&
+
\dfrac{2}{\phi'} \left(\frac{A'}{A} + \frac{C'}{C} + \frac{X'}{X} \right)
X G_{4\phi} 
+
B X \left( - \frac{A'^2}{A^2}- \frac{C'^2}{C^2}+\frac{A'B'}{AB}+\dfrac{A'C'}{AC}+\dfrac{B'C'}{BC} +\dfrac{2A''}{A}+\dfrac{2C''}{C} \right) 
\left( G_{4X}-\frac{1}{2}G_{5\phi} \right)
\nonumber\\&&
+
BX' \left( \frac{C'}{C}+\frac{A'}{A} \right) \left( G_{4X}-G_{5\phi} \right)+4X G_{4\phi\phi}
+ 2B\phi' \left( \frac{C'}{C} + \frac{A'}{A} - \frac{X'}{X} \right) X  G_{4\phi X}
\nonumber\\&&
+ 2B \left( \frac{C'}{C}+\frac{A'}{A} \right) XX' G_{4XX}
-
\frac{B^2\phi'}{2C} \left[\frac{A''C'}{A}-\frac{A'^2C'}{2A^2} + \frac{A'}{A} \left(C''-\dfrac{C'^2}{2C}+\frac{B'C'}{B}+\frac{3C'X'}{2X} \right) \right] X G_{5X}
\nonumber\\&&
-B\phi' X \left( \frac{C'}{C}+\frac{A'}{A} \right) G_{5\phi \phi}
+
B \left[\dfrac{A'C'}{AC}-\frac{X'}{X} \left( \frac{C'}{C}+\frac{A'}{A} \right) \right] X^2
G_{5\phi X}
\nonumber\\&&
-\frac{A'B^2C' \phi'}{2AC} XX' G_{5XX},
\end{eqnarray}
\end{widetext}

%---------   References   ---------%

\end{document}